\pdfoutput=1
\documentclass[aps, reprint, superscriptaddress,nofootinbib, showpacs,preprintnumbers, floatfix]{revtex4-1}
\usepackage{amsmath, latexsym, amssymb, hyperref, graphicx, color, multirow,slashed}
\usepackage[table]{xcolor}
\usepackage{tabularx}
\usepackage[T1]{fontenc}
\usepackage[capitalize]{cleveref}
\definecolor{nicered}{rgb}{.7,.1,.1}
\definecolor{nicegreen}{rgb}{.1,.5,.1}
\definecolor{darkblue}{rgb}{0,0,.5}
\hypersetup{colorlinks, citecolor=nicegreen,linkcolor=nicered, urlcolor=darkblue}
\usepackage{caption}
\usepackage{subcaption}
\usepackage[normalem]{ulem}

\newcommand{\gev}{~\text{GeV}}
\newcommand{\tev}{~\text{TeV}}
\newcommand{\mev}{~\text{MeV}}

\newcommand{\onbb}{0\nu\beta\beta}

\newcommand{\gsim}{\buildrel > \over {_\sim}}
\def\nn{\nonumber}

\begin{document}

\preprint{ACFI-T20-11}


\title{Left-right symmetry and leading contributions to neutrinoless double beta decay}
\author{Gang Li}
\email{ligang@umass.edu}
\affiliation{Amherst Center for Fundamental Interactions, Department of Physics, University of Massachusetts, Amherst, MA 01003, USA.}
\author{Michael Ramsey-Musolf}
\email{mjrm@sjtu.edu.cn,\, mjrm@physics.umass.edu}
\affiliation{Tsung-Dao Lee Institute and School of Physics and Astronomy, Shanghai Jiao Tong University, 800 Dongchuan Road, Shanghai, 200240 China.}
\affiliation{Amherst Center for Fundamental Interactions, Department of Physics, University of Massachusetts, Amherst, MA 01003, USA.}
\affiliation{Kellogg Radiation Laboratory, California Institute of Technology, Pasadena, CA 91125 USA.}
\author{Juan Carlos Vasquez}
\email{jvasquezcarm@umass.edu}
\affiliation{Amherst Center for Fundamental Interactions, Department of Physics, University of Massachusetts, Amherst, MA 01003, USA.}


\begin{abstract}
\noindent
We study the impact   of the mixing  (LR mixing)   between the standard model $W$ boson and its hypothetical,  heavier right-handed parter $W_R$ on the neutrinoless double beta decay ($\onbb$-decay) rate. Our study is done in the minimal left-right symmetric model assuming type-II dominance scenario with charge conjugation as  the left-right symmetry. We then show that  the $\onbb$-decay rate may be dominated by the contribution proportional to this LR mixing, which at the hadronic level induces the leading-order contribution to  the interaction between two pions and two charged leptons. The resulting long-range pion exchange contribution can significantly enhance the decay rate compared to previously considered short-range contributions. Finally, we find that  even if future cosmological experiments rule out the inverted hierarchy for neutrino masses,  there are still good prospects for a positive signal in the next generation of  $\onbb$-decay experiments. 
\end{abstract}
\pacs{}

\maketitle
%
%
Determining the properties of the light neutrinos under charge conjugation is a key challenge for particle and nuclear physics. As the only electrically neutral fermions in the Standard Model (SM) of particle physics, neutrinos are the sole SM candidates for possessing a Majorana mass. The corresponding term in the Lagrangian breaks the conservation of total lepton number ($L$) by two units:  $\mathcal{L}_M\supset - y_\nu \overline {\ell^C} H^T H \ell /\Lambda$, where  $\ell$ and $H$ are the SM left handed lepton doublet and Higgs doublet, respectively, and $\Lambda$ is a mass scale whose presence is needed to maintain dimensionality. After the neutral component of the Higgs doublet obtains a vacuum expectation value (vev) $v/\sqrt{2}$ , the resulting Majorana mass operator is $\mathcal{L}_M \to - (m_\nu/2) \overline{\nu^c} \nu$, with $m_\nu = y_\nu v^2/\Lambda$. For $y_\nu \sim \mathcal{O}(1)$, the observed scale of light neutrino masses consistent with oscillation experiments~\cite{Tanabashi:2018oca} and cosmological bounds~\cite{Capozzi:2020qhw,Capozzi:2017ipn} would imply $\Lambda\gsim 10^{15}$ GeV.

An experimental determination that neutrinos are Majorana fermions could, thus, provide circumstantial evidence for $L$-violating processes at ultra-high energy scales involving new particles not directly accessible in the laboratory. In the widely-considered see-saw mechanism, the $L$-violating, out-of-equilibrium decays of these particles (fermions) could generate the cosmic matter-antimatter asymmetry~\cite{Fukugita:1986hr}. Neutrino oscillation experiments are agnostic regarding the existence of a Majorana neutrino mass term. However, the observation of $\onbb$-decay in the nuclear transition $(A,Z)\to (A,Z+2) +e^- +e^-$~\cite{Furry:1939qr} -- a process that also violates $L$ by two units -- would provide conclusive evidence that light neutrinos are Majorana fermions~\cite{Schechter:1981bd}.  

The recent $\onbb$-decay search in the KamLAND-Zen experiment~\cite{KamLAND-Zen:2016pfg} provides the most stringent upper limit on the effective Majorana mass $|m_{\beta\beta}|$, which is $0.061-0.165$~eV at 90\% confidence level (C.L.), where the range reflects the uncertainty in nuclear matrix element (NME) computations. In the three-neutrino framework~\cite{Fogli:2005cq}, $|m_{\beta\beta}|$ 
depends on the neutrino mass spectrum. In the inverted hierarchy (IH) it is bounded below $|m_{\beta\beta}| \gsim 0.01$ eV, while in the normal hierarchy (NH) it can be vanishingly small. The next generation of  $\onbb$-decay searches with ton-scale detectors~\cite{Albert:2017hjq,Kharusi:2018eqi,Abgrall:2017syy,Armengaud:2019loe,CUPIDInterestGroup:2019inu,Paton:2019kgy} aim for sensitivities for $|m_{\beta\beta}|$ as low as $ ~ 0.01$ eV. If neutrinos are Majorana fermions, and if the IH is realized in nature, one would thus expect a non-zero result in the ton-scale experiments.

Cosmological observations provide complementary information on neutrino masses, currently constraining the sum of neutrino masses (dubbed $\Sigma m_\nu$) to be smaller than 0.12~eV at the $2\sigma$ level~\cite{Aghanim:2018eyx}. 
Global fits~\cite{Capozzi:2017ipn,Capozzi:2020qhw} of neutrino oscillation data, $\onbb$-decay search results, and cosmological surveys show that the NH is favored over the IH at about $2\sigma$ level. For future cosmological surveys~\cite{Brinckmann:2018owf,Abazajian:2019eic,Hazumi:2012gjy,Levi:2019ggs,Scaramella:2015rra}, it is possible to exclude the IH, while the favored $|m_{\beta\beta}|$ may be out the reach of ton-scale $\onbb$-decay experiments~\cite{Albert:2017hjq,Kharusi:2018eqi,Abgrall:2017syy,Armengaud:2019loe,CUPIDInterestGroup:2019inu,Paton:2019kgy} in the three-neutrino framework. Then, it is natural to ask how one could interpret a $\onbb$-decay signal if cosmological measurements and/or future  oscillation experiments demonstrate conclusively that the light neutrino mass ordering is in the NH.

\begin{figure}
\captionsetup[subfigure]{justification=centering}
     \centering
     \begin{subfigure}[b]{0.2\textwidth}
         \centering
         \includegraphics[width=\textwidth]{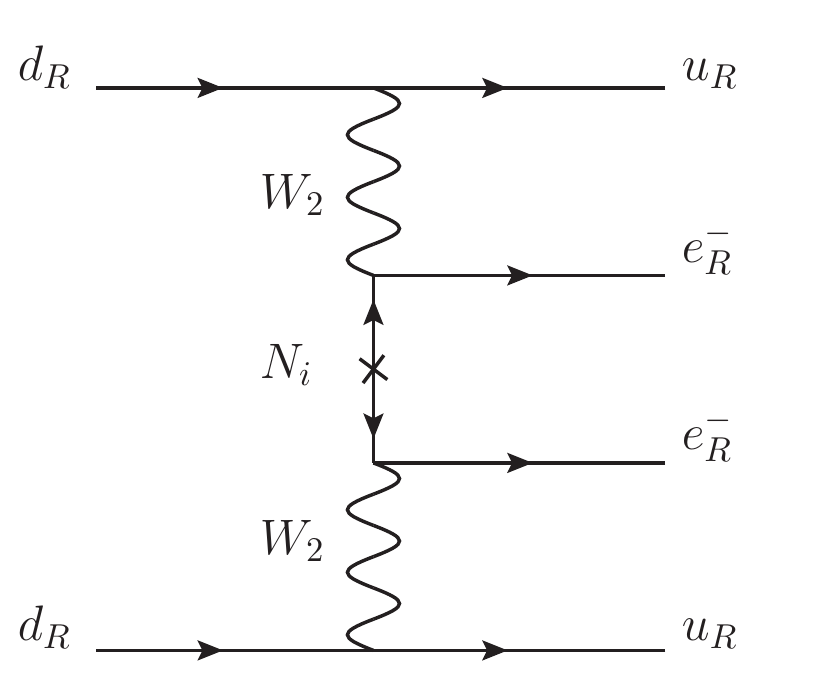}
         \caption{ }
     \end{subfigure}
     \hfill
     \begin{subfigure}[b]{0.2\textwidth}
         \centering
         \includegraphics[width=\textwidth]{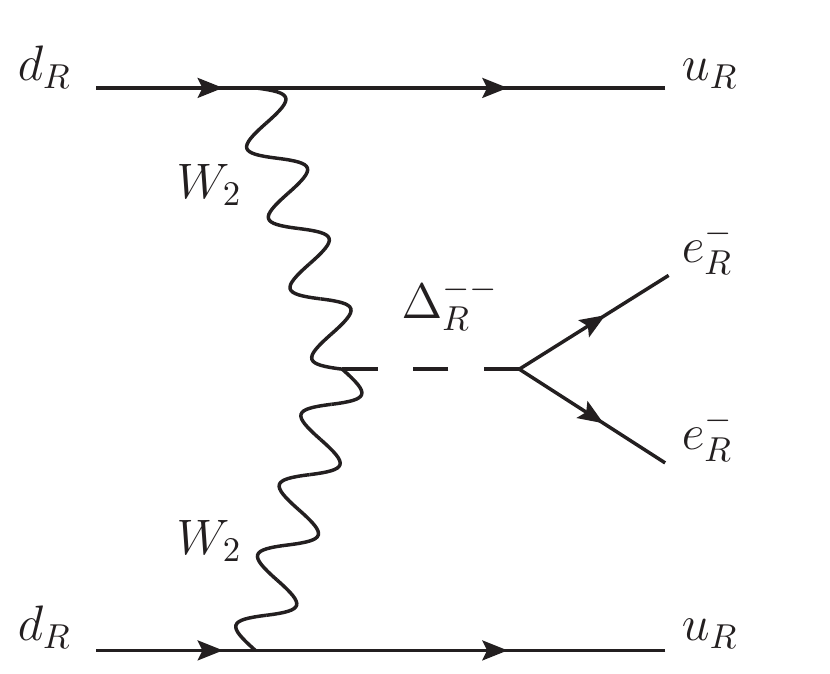}
         \caption{ }
     \end{subfigure}
     \hfill
     \begin{subfigure}[b]{0.2\textwidth}
         \centering
         \includegraphics[width=\textwidth]{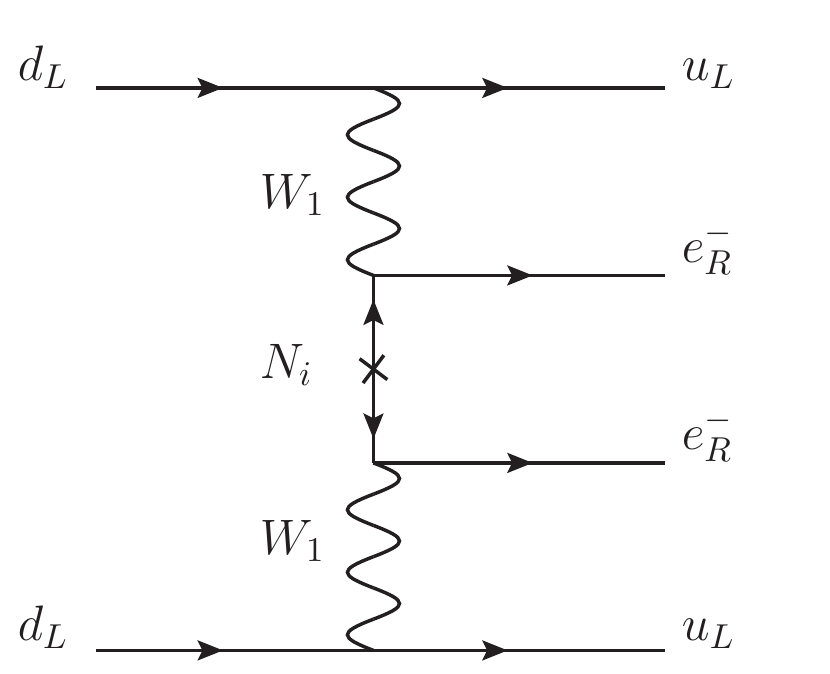}
         \caption{ }
     \end{subfigure}
     \hfill
     \begin{subfigure}[b]{0.2\textwidth}
         \centering
         \includegraphics[width=\textwidth]{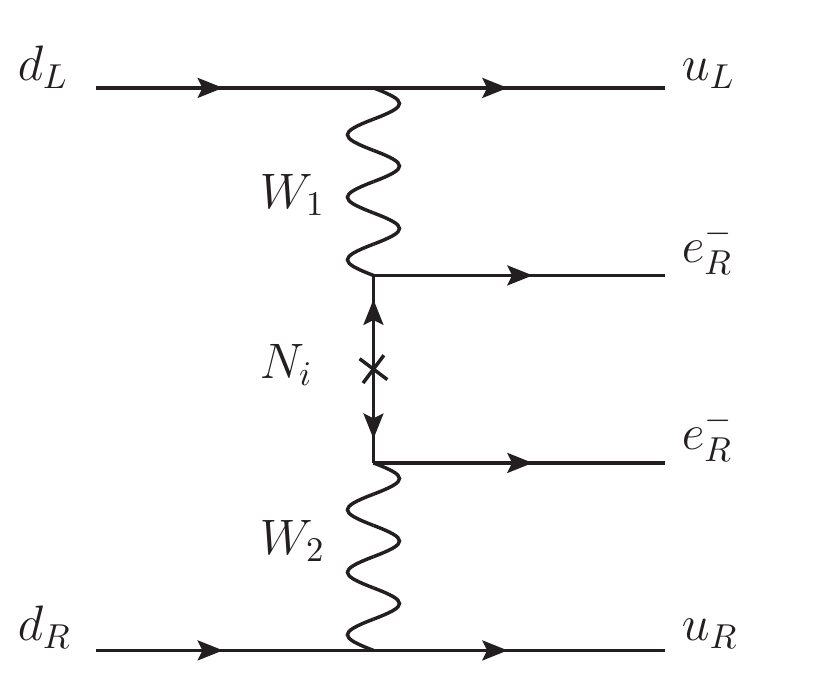}
         \caption{ }
     \end{subfigure}
        \caption{Feynman diagrams in the mLRSM contributing to the $\onbb$-decay.}
        \label{fig:diagrams}
\end{figure}

Here, we address this question in the context of one of the most extensively studied extensions of the SM that generically implies the existence of Majorana neutrinos: the  minimal left-right symmetric model (mLRSM)~\cite{Pati:1974yy,Mohapatra:1974gc,Senjanovic:1975rk,SENJANOVIC1979334,Mohapatra:1979ia,Mohapatra:1980yp}.  This model may have TeV scale  new particles and the contributions to the $\onbb$-decay from the new right-handed sector can be appreciable. The light neutrino and new physics contributions are characterized by $G_F^2 |m_{\beta\beta}|/p^2$ and $c/\Lambda^5$~\cite{Mohapatra:1998ye,Prezeau:2004cs,Cirigliano:2004tc,Rodejohann:2011mu}, respectively. Here, the virtual neutrino momentum $p\simeq 100~\text{MeV}$, $G_F$ is the Fermi constant and $c$ denotes new Yukawa and/or gauge couplings. For $c\simeq \mathcal{O}(1)$ and $|m_{\beta\beta}|\simeq 0.1~(0.01)$~eV, the new physics contribution can be comparable to the light neutrino contribution if $\Lambda\simeq 3.7~(5.9)~\text{TeV}$. In particular, it has been shown~\cite{Tello:2010am} that in the mLRSM the contributions coming from heavy neutrinos from the exchange of two right-handed $W_R$ bosons (the RR amplitude), see Fig.~\ref{fig:diagrams}(a), are sizable. Nonetheless, the bulk of the mLRSM parameter space would remain largely inaccessible to ton-scale $\onbb$-decay searches if cosmological data push the bound on $\Sigma m_\nu$ below $\sim 0.1$ eV.

In what follows, we show that this conclusion changes dramatically in the presence of mixing between the left- and right-handed gauge bosons. This mixing results in contributions to the decay amplitude involving the exchange of heavy right-handed neutrinos,  one SM $W$ boson (predominantly left-handed) and one heavy $W$ boson (predominantly right-handed) -- a contribution we denote as the LR amplitude, see Fig.~\ref{fig:diagrams}(d). In Ref.~\cite{Chakrabortty:2012mh,Barry:2013xxa} it was found that the LR amplitude is suppressed with respect to the RR amplitude due to the upper bounds on the $W_L$-$W_R$ mixing angle. However,  those studies did not include long-range contributions associated with pion exchange that significantly enhance LR amplitude and  can compensate for  this  suppression~\cite{Prezeau:2003xn}.  In this Letter, we compute these long-range contributions using state-of-the art information on hadronic and nuclear matrix elements as well as phenomenological constraints on the relevant mLRSM parameters. We find that even in the presence of prospective, stringent cosmological bounds on $\Sigma m_\nu$ and possible exclusion of the IH, there exists ample opportunity for the observation of a signal in next generation $\onbb$-decay searches.

This framework entails extending the SM gauge group to $SU(3)_C\times SU(2)_L\times SU(2)_R \times  U(1)_{B-L}$, where $B$ and $L$ denote the SM abelian baryon and lepton quantum numbers. The Higgs sector consists of two scalar triplets $\Delta_L\in (1,3,2)$, $\Delta_R\in(3,1,2)$ and one bidoublet $\Phi\in(2,2,0)$, where $(X,Y,Z)$ denote the representations under the SU(2)$_{R,L}$ and U(1)$_{B-L}$ groups. The neutral components of the bidoublet field $\Phi$ obtain vevs: $\langle \Phi \rangle \to \mathrm{diag}\, \{v_1, v_2 e^{i \alpha}\}$ with $v=\sqrt{v_1^2+v_2^2}$ and $\alpha$ being the spontaneous CP-violating phase.

Of particular relevance to $\onbb$-decay is the charged-current Lagrangian
   \begin{align} 
   \mathcal{L}_{CC} &=
   -\frac{g}{\sqrt{2}}\, \sum_{A=L,R} \Bigl\{ \bar{u}_{A i} V^{\text{CKM}}_{Aij}\slashed{W}_A d_{A j}\\
   \nonumber
  &- \bar{e}_{Ai} V_{Aij} \slashed{W}_{A} \nu_{Aj} \Bigr\}+  \text { h.c.}\;,
   \end{align}
where $A=L,R$ and $V_{L,R}^{\text{CKM}}$ and $V_{L,R}$ are the Cabibo-Kobayashi-Maskawa (CKM) and lepton-mixing matrices, respectively. The $L,R$ gauge bosons in terms of the light and heavy mass eigenstates  $W_1$ and $W_2$ are given by  $W_{L,R}^{+\mu}=\cos\xi W_{1,2}^{+\mu}\mp\sin\xi e^{\mp i\alpha} W_{2,1}^{+\mu}$ where  $\tan\xi =  \lambda \sin(2\beta) e^{i\alpha}$ with  $\tan\beta = v_2/v_1$ and $\lambda = M_{W_1}^2/M_{W_2}^2$. 

Direct searches of the $W_R$ boson require $M_{W_R}\simeq M_{W_2}>4.4\tev$~\cite{Sirunyan:2018pom},
implying $\lambda <3.4\times 10^{-4}$. Tests of CKM unitarity place constraints on $\xi$. From recent results for the radiative corrections to nuclear $\beta$-decay~\cite{Seng:2018yzq,Blucher:2008zz}, $0.25\times 10^{-3}\leq \xi \leq 1.25\times 10^{-3}$ is allowed at 95\% C.L. in order to restore the CKM unitarity. We will consider the range $0\leq \xi \leq 1.25\times 10^{-3}$.
If the LR symmetry is taken to be parity ($\mathcal{P}$), $|\sin\alpha \tan(2\beta)|<2m_b/m_t$~\cite{Senjanovic:2014pva,Senjanovic:2015yea} with $m_b$ and $m_t$ being the bottom and top quark masses, respectively. 
No such constraint exists when charge conjugation ($\mathcal{C}$) is the LR symmetry~\cite{Senjanovic:2014pva,Senjanovic:2015yea}. In Ref.~\cite{Cirigliano:2018yza}, they also consider the LR symmetry as $\mathcal{C}$ but with the maximum of $\tan\beta$ being $m_b/m_t$.
Constraints from kaon CP violation and neutron electric dipole moments apply when $\alpha\not= 0$~\cite{Zhang:2007da,Maiezza:2010ic,Engel:2013lsa,Bertolini:2014sua,Cirigliano:2016yhc,Bertolini:2019out}. Here we consider  $\mathcal{C}$ as the LR symmetry and  $\alpha=0$ since our results are rather insensitive to fundamental sources of CP violation.
There is no direct experiment bound on $\tan\beta$ so we choose $\tan\beta<0.5$ to keep the bidoublet Yukawa coupling of order unity.  We will assume  $M_{W_R} = 7$~TeV, which satisfies all the aforementioned constraints as well as the requirement of $M_{W_R}\gsim 6$ TeV from the renormalization group evolution (RGE) analysis~\cite{Maiezza:2016ybz}.

For purposes of illustration, we follow Ref.~\cite{Tello:2010am} and assume  \lq\lq type-II dominance\rq\rq\, for neutrino masses~\footnote{The type-I seesaw scenario was studied in Ref.~\cite{Cirigliano:2018yza} where the new physics contribution can also dominate over standard light neutrino exchange scenario. }. In this scenario, $m_{N_i}\propto m_{\nu_i}$, one has $V_L = V_R^*$~\cite{Tello:2010am}.
Using the light neutrino mass difference from solar and atmospheric neutrinos~\cite{deSalas:2020pgw},  $M_{W_R}=gv_R$, and fixing the neutrino mass $m_{N_{\text{max}}}$ ($=m_{N_3}$ for the NH and $=m_{N_2}$ for the IH), it is possible to obtain all the neutrino masses in terms of the lightest neutrino mass $m_{\nu_{\text{min}}}$.

The  effective Lagrangian below the electroweak scale is
\begin{align}
\label{Leff}
\mathcal{L}_{\text{eff}}&=\dfrac{G_F^2}{\Lambda_{\beta\beta}}\big[ C_{3R} (\mathcal{O}_{3+}^{++}-\mathcal{O}_{3-}^{++})(\bar{e}e^c -\bar{e}\gamma_5 e^c)\nn\\
&+C_{3L} (\mathcal{O}_{3+}^{++}+\mathcal{O}_{3-}^{++})(\bar{e}e^c -\bar{e}\gamma_5 e^c)\\
&+C_{1} \mathcal{O}_{1+}^{++}(\bar{e}e^c -\bar{e}\gamma_5 e^c)
+C_{1}^\prime \mathcal{O}_{1+}^{++\prime}(\bar{e}e^c -\bar{e}\gamma_5 e^c)\big]\;, \nn
\end{align}
where~\cite{Prezeau:2003xn}
\begin{align}
\mathcal{O}_{3\pm}^{++}=&(\bar{q}^\alpha_L \tau^+ \gamma^{\mu} q^\alpha_L)(\bar{q}^\beta_L \tau^+ \gamma_{\mu} q^\beta_L) \pm \, (L\to R)\;, \\
\mathcal{O}_{1+}^{++}=&(\bar{q}^\alpha_L \tau^+ \gamma^{\mu} q^\alpha_L)(\bar{q}^\beta_R \tau^+ \gamma_{\mu} q^\beta_R)\;,\\
\mathcal{O}_{1+}^{++^\prime}=&(\bar{q}^\alpha_L \tau^+ \gamma^{\mu} q^\beta_L)(\bar{q}^\beta_R \tau^+ \gamma_{\mu} q^\alpha_R)\;,
\end{align}
and $\alpha,\beta$ are the color indices, $\tau^{\pm}= (\tau^1\pm\tau^2)/2$, $\tau^1$ and $\tau^2$ are the Pauli matrices. 

Wilson coefficients $C_{3R}$, $C_{3L}$ and $C_1$ are obtained by integrating out the $W_{1,2}$ and $N_i$ arising respectively from the amplitudes in Fig.~\ref{fig:diagrams}(a)(c)(d).  We evolve them from the scale $\mu=M_{W_2}$ to an appropriately chosen hadronic scale $\Lambda_\mathrm{H}=2$ GeV~\cite{Aoki:2019cca}. The RGE proceeds in two steps: (a) $\mu= M_{W_2}\to M_{W_1}$; (b) $\mu=M_{W_1}\to\Lambda_\mathrm{H}$ and it gives~\cite{Cirigliano:2018yza,Cirigliano:2017djv,Liao:2019gex} 
\begin{subequations}
\label{eq:wilson2}
\begin{align}
\begin{pmatrix}
C_1 (\Lambda_H) \\
C_1^\prime (\Lambda_H)
\end{pmatrix}
&= 
\begin{pmatrix}
0.90 & 0\\
0.48 & 2.32
\end{pmatrix}
\begin{pmatrix}
C_1(M_{W_1})\\
C_1^{\prime}(M_{W_1})
\end{pmatrix}\;,\\
C_{3L}(\Lambda_H)&= 0.81 C_{3L}(M_{W_1})\;,\\
C_{3R}(\Lambda_H)&= 0.71 C_{3R}(M_{W_2})\;,
\end{align}
\end{subequations}
where $C_1^{\prime}(M_{W_1})=0$ and it appears due to the RGE of $C_1$. In Eq.~\eqref{eq:wilson2}, the non-vanishing Wilson coeffcients at the electroweak scale are given by $C_1(M_{W_1})=-4\lambda\xi$, $C_{3L}(M_{W_1})=\xi^2$ and $C_{3R}(M_{W_2})=\lambda^2( 1+4\Lambda_{\beta\beta}^2/M_{\Delta_R}^2)$ with $1/\Lambda_{\beta\beta}=\sum_{i=1}^{3}|V_{Rei}|^2/m_{N_i}$. 
Note that $\mathcal{O}_{3L}\equiv \mathcal{O}_{3+}^{++}+\mathcal{O}_{3-}^{++}$ and $\mathcal{O}_{1+}^{++}$ are matched to effective operators above the electroweak scale, which however do not evolve under QCD running~\cite{Cirigliano:2018yza}, so that the RGE only includes   step $(b)$. 


The doubly charged scalar, depicted in Fig.~\ref{fig:diagrams}(b), contributes solely to $C_{3R}$. 
When the LR symmetry holds,  this contribution is negligible due to collider bounds~\cite{Aaboud:2017qph} and charged lepton flavor violation constraints~\cite{Tello:2010am}.  On the contrary, when the LR symmetry is explicitly broken, these constraints are relaxed and the corresponding contribution to the $\onbb$-decay rate can be appreciable. For a discussion, and the possible interplay with prospective future low- and high-energy probes, see Ref.~\cite{Dev:2018sel}.
Here, we assume a LR-symmetric Lagrangian and leave the analysis of the interesting case when it is broken for a future work. 

We now map the operators in Eq.~(\ref{Leff}) at GeV scale $\sim \Lambda_H$ onto an effective hadron-lepton Lagrangian below that scale~\cite{Prezeau:2003xn,Graesser:2016bpz,Cirigliano:2018yza}  using chiral perturbation theory ($\chi$PT)~\cite{Bernard:1995dp,Gasser:1983yg}. Matching entails identifying all operators at a given chiral order  that transform under chiral SU(2) the same way as the four-quark factor of a given operator in Eq.~(\ref{Leff})~\cite{Prezeau:2003xn,Kaplan:1992vj}. We refer the reader to Ref.~\cite{Prezeau:2003xn} for a detailed derivation, and here simply quote the results. 

The hadron-lepton Lagrangian for the $\pi\pi \bar{e}e^c$, $\bar{N}N\pi \bar{e}e^c $ and $\bar{N}N\bar{N}N\bar{e}e^c$ operators up to NNLO in chiral expansion is~\cite{Prezeau:2003xn}
\begin{align}
\label{eq:hadron-lepton}
\mathcal{L}_{\chi\text{PT}}=&\dfrac{G_F^2F_\pi^2}{\Lambda_{\beta\beta}}\Big\{\Lambda_\chi^2 \pi^-\pi^- \bar{e}(\beta_1+\beta_2\gamma^5)e^c\nn\\
&+\partial_\mu \pi^-\partial^\mu \pi^- \bar{e}(\beta_3+\beta_4\gamma^5)e^c\nn\\
&+\Lambda_\chi/F_\pi \bar{N}i\gamma_5 \tau^+ \pi^- N  \bar{e}(\zeta_5+\zeta_6\gamma^5)e^c\nn\\
&+1/F_\pi^2 \bar{N}\tau^+N \bar{N}\tau^+N\bar{e}(\xi_1+\xi_4\gamma_5)e^c\nn\\
&+\text{h.c.}\Big\}\;.
\end{align}

The first two-pion term contributes to the amplitude $\mathcal{A}(nn\to ppe^-e^-)$ at order of $p^{-2}$ with $p\lesssim m_\pi$ being the typical momentum transfer. When this leading-order (LO) amplitude $\mathcal{A}^{\text{LO}}$ is present as in the mLRSM, it can give a dominant long-range contribution to the half-life of $\onbb$-decay~\cite{Prezeau:2003xn}. The one-pion and four-nucleon and another two-pion terms, however, contribute at next-to-next-to LO (NNLO) to the amplitude $\mathcal{A}^{\text{NNLO}}\sim p^0$. The dimensionless coefficients are expressed as~\cite{Prezeau:2003xn} $\beta_1=-\beta_2=\ell_1^{\pi\pi} C_1 +\ell_1^{\pi\pi\prime} C_1^\prime$, $\beta_3=-\beta_4=\ell_3^{\pi\pi} (C_{3L} + C_{3R})$, $\zeta_5=-\zeta_6=\ell_3^{\pi N} (C_{3L}+C_{3R})$, and
$\xi_{1}=-\xi_{4}=\ell_{1}^{NN} C_1 +\ell_{1}^{NN\prime} C_1^\prime +\ell_{3}^{NN} (C_{3L}+C_{3R})$. Furthermore, $g_A=1.271$,   $\Lambda_\chi=4\pi F_{\pi}$ with $F_\pi=92.28\mev$ and $\ell_i$ are the low energy constants (LECs).  Using the lattice calculations~\cite{Nicholson:2018mwc}, we get $\ell_1^{\pi\pi}=-(0.71\pm 0.07)$, $\ell_1^{\pi\pi\prime} =-(2.98\pm 0.22)$ and $\ell_3^{\pi\pi}=0.60\pm 0.03$ in the modified minimal substraction ($\overline{\text{MS}}$) scheme at $\mu=2\gev$~\cite{Cirigliano:2018yza}. The LECs for $\bar{N}N\pi \bar{e}e^c$ and $\bar{N}N\bar{N}N \bar{e}e^c$ interactions are unknown and are estimated using the naive dimensional analysis (NDA)~\cite{Manohar:1983md} with $\ell_3^{\pi N}\sim \mathcal{O}(1)$ and $\ell_{1}^{NN},\ell_{1}^{NN\prime}, \ell_{3}^{NN}\sim \mathcal{O}(1)$. 

  The four-nucleon interaction in Eq.~\eqref{eq:hadron-lepton} merits a more detailed discussion.  In Ref.~\cite{Cirigliano:2018yza},  it was observed that a  consistent renormalization of  the amplitude induced by the operators $\mathcal{O}_{1+}^{++}\,,\mathcal{O}_{1+}^{++'}$ 
requires inclusion of a LO four-nucleon counterterm~\cite{Cirigliano:2018hja}. While its presence does not change the magnitude of the NNLO contributions  (barring accidental cancellations), it does introduce additional hadronic uncertainties at LO. To check how this new source of uncertainty might affect our results, we have taken the natural assumption  that this new contribution gives an additional  $100\%$  contribution to the decay rate and found that our conclusions  remain the same. Finally and notwithstanding the above arguments, the uncertainty might be bigger, as suggested by the RGE analysis of Ref.~\cite{Cirigliano:2018yza}. However, this issue is still far from settled until the finite piece of the LO four-nucleon    counterterm is taken from a more reliable source, such as lattice QCD for instance.        
\begin{figure}
\centering
\includegraphics[width=1\columnwidth]{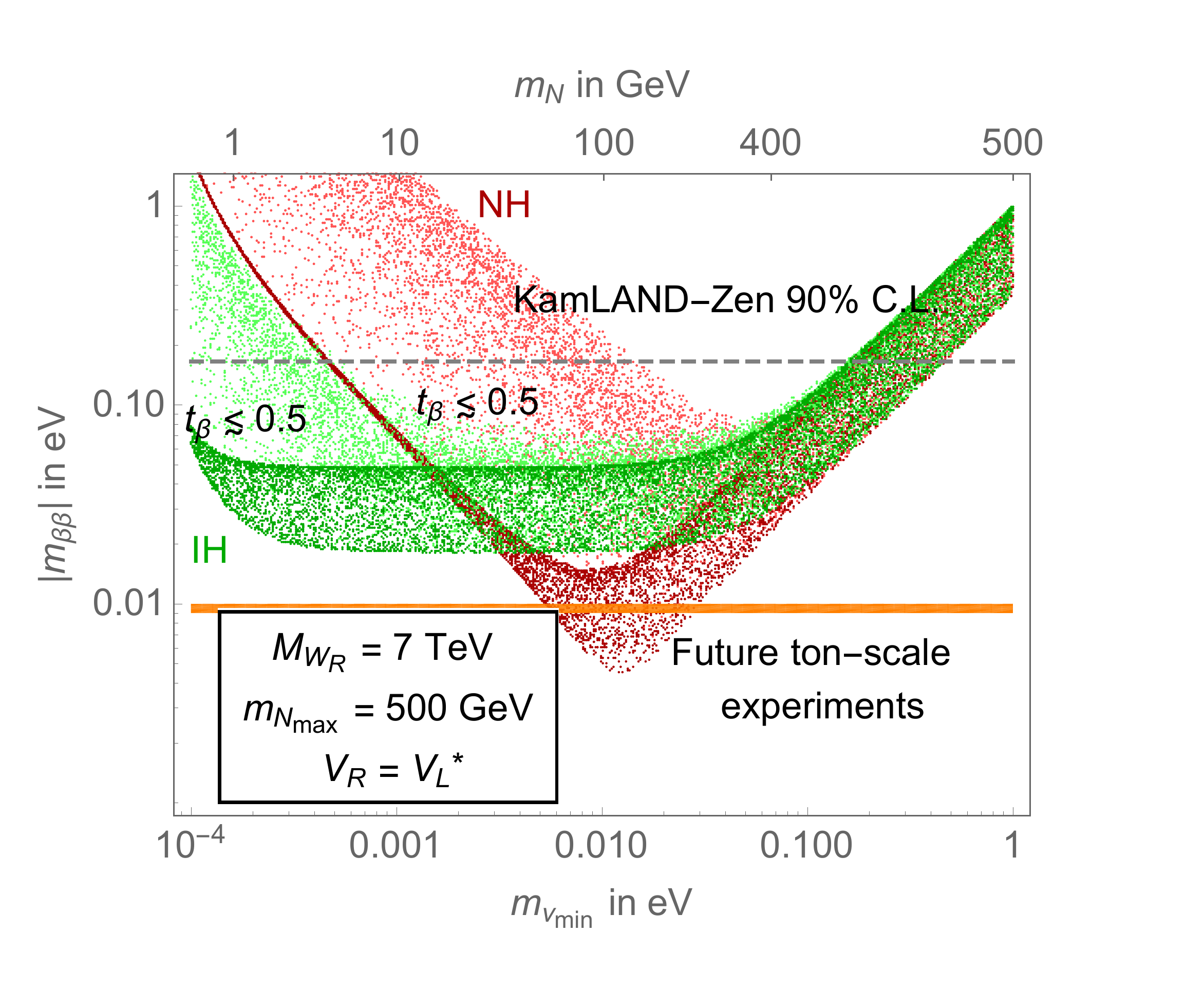}
\caption{ Effective Majorana mass as a function of the lightest neutrino mass. 
The central values of the mixing angles and the Dirac CP-violating phase in $V_L$ are quoted from Ref.~\cite{deSalas:2020pgw},
and the Majorana phases are marginalized.
The allowed regions with $\tan\beta=0$ and $\tan\beta\leq 0.5$ are depicted in darker and lighter colors. Red (green) dots denote the NH (IH) of neutrino mass ordering. Gray and orange lines represent the current and expected limits  limits from the KamLAND-Zen~\cite{KamLAND-Zen:2016pfg} and future ton-scale experiment~\cite{Albert:2017hjq,Kharusi:2018eqi}, respectively. The lightest heavy neutrino mass is also given in the upper horizontal axis. }
\label{decayrates_mvmin}
\end{figure}
From Eq.~\eqref{eq:hadron-lepton}, we obtain the decay half-life
\begin{align}
\label{eq:half-life}
  (  T^{0\nu}_{1/2})^{-1}= &\, G_{0\nu} \cdot\mathcal{M}_{\nu}^{2}\,|m_{\beta\beta}|^2\nonumber \\ =  &\, G_{0\nu} \cdot\mathcal{M}_{\nu}^{2}\left(\left|m_{\nu}^{ee}\right|^{2}+\left|m_N^{ee}\right|^2\right)\;,
\end{align}
where
\begin{align}
m_{\nu}^{ee}   &  \simeq \sum_{i=1}^{3} |V_{Lei}|^2 m_{\nu_i}(1+\ell_{\nu}^{NN} \delta^{\nu}_{NN})\;,
\end{align}
and
\begin{align}
\label{mNeff}
&|m_{N}^{ee}|^2  =  \dfrac{\Lambda_\chi^4}{72\Lambda_{\beta\beta}^2}\dfrac{\mathcal{M}_0^2}{\mathcal{M}_\nu^2}\times\bigg[(\beta_1-\zeta_5\delta_{N\pi}-\beta_3\delta_{\pi\pi}+\xi_1\delta_{NN} )^2\nn\\
&\quad +(\beta_2-\zeta_6\delta_{N\pi}-\beta_4\delta_{\pi\pi}+\xi_4\delta_{NN}  )^2\bigg]
\end{align}
with $m_N=939\mev$ and
\begin{align}
\delta_{\pi\pi}&= \dfrac{2m_\pi^2}{\Lambda_\chi^2}\dfrac{\mathcal{M}_2}{\mathcal{M}_0}, &
\delta_{N\pi}&= \dfrac{\sqrt{2}m_\pi^2}{g_A \Lambda_\chi m_N}\dfrac{\mathcal{M}_1}{\mathcal{M}_0}, \\
\delta_{NN}^{\nu}&=\dfrac{2m_\pi^2 }{g_A^2 \Lambda_\chi^2} \dfrac{\mathcal{M}_{NN}}{\mathcal{M}_\nu}, &
\delta_{NN}&=\dfrac{12m_\pi^2}{g_A^2 \Lambda_\chi^2}\dfrac{\mathcal{M}_{NN}}{\mathcal{M}_0}.
\end{align}

\begin{figure}
\centering
\includegraphics[width=.9\columnwidth]{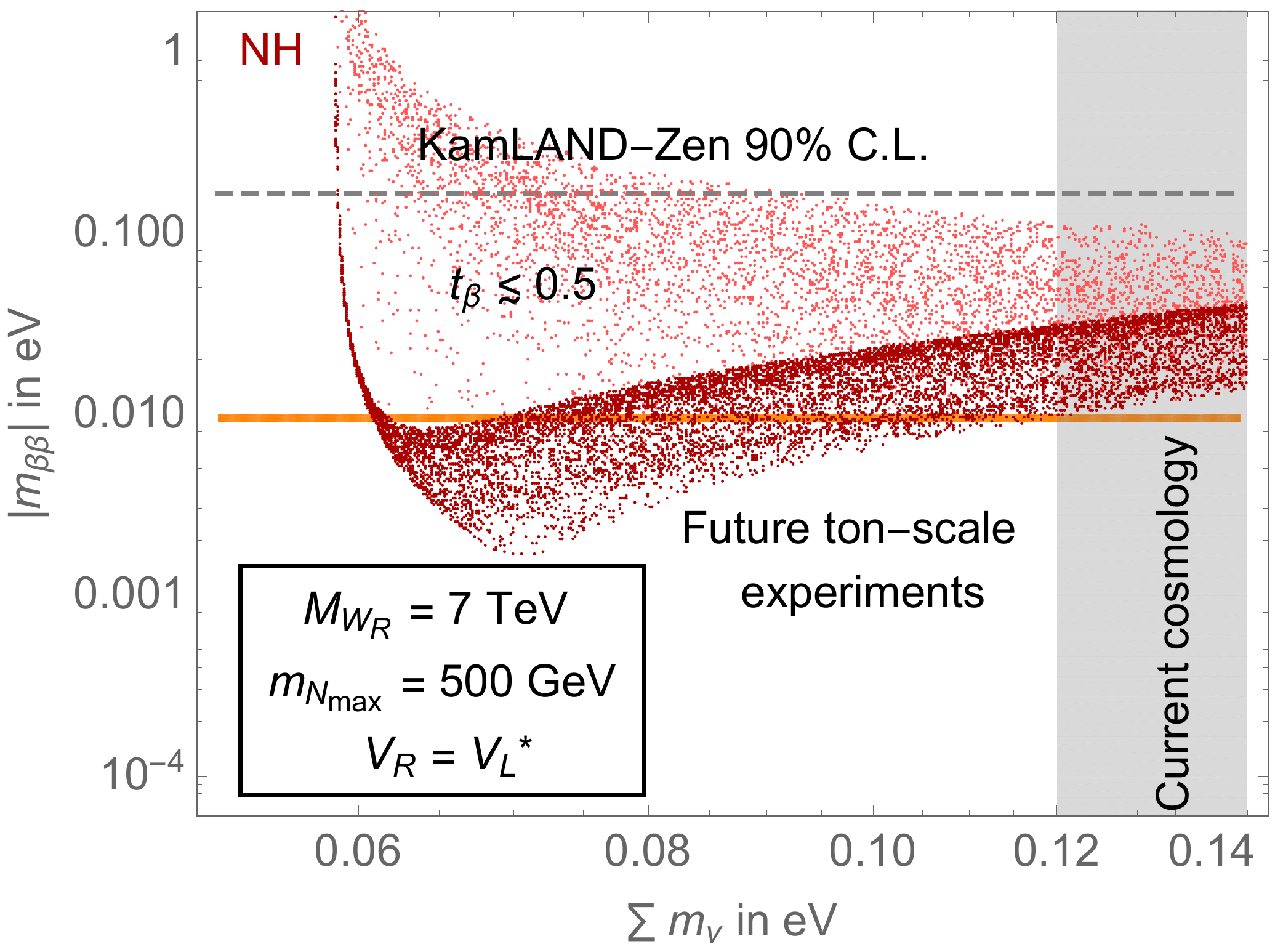}
\includegraphics[width=.9\columnwidth]{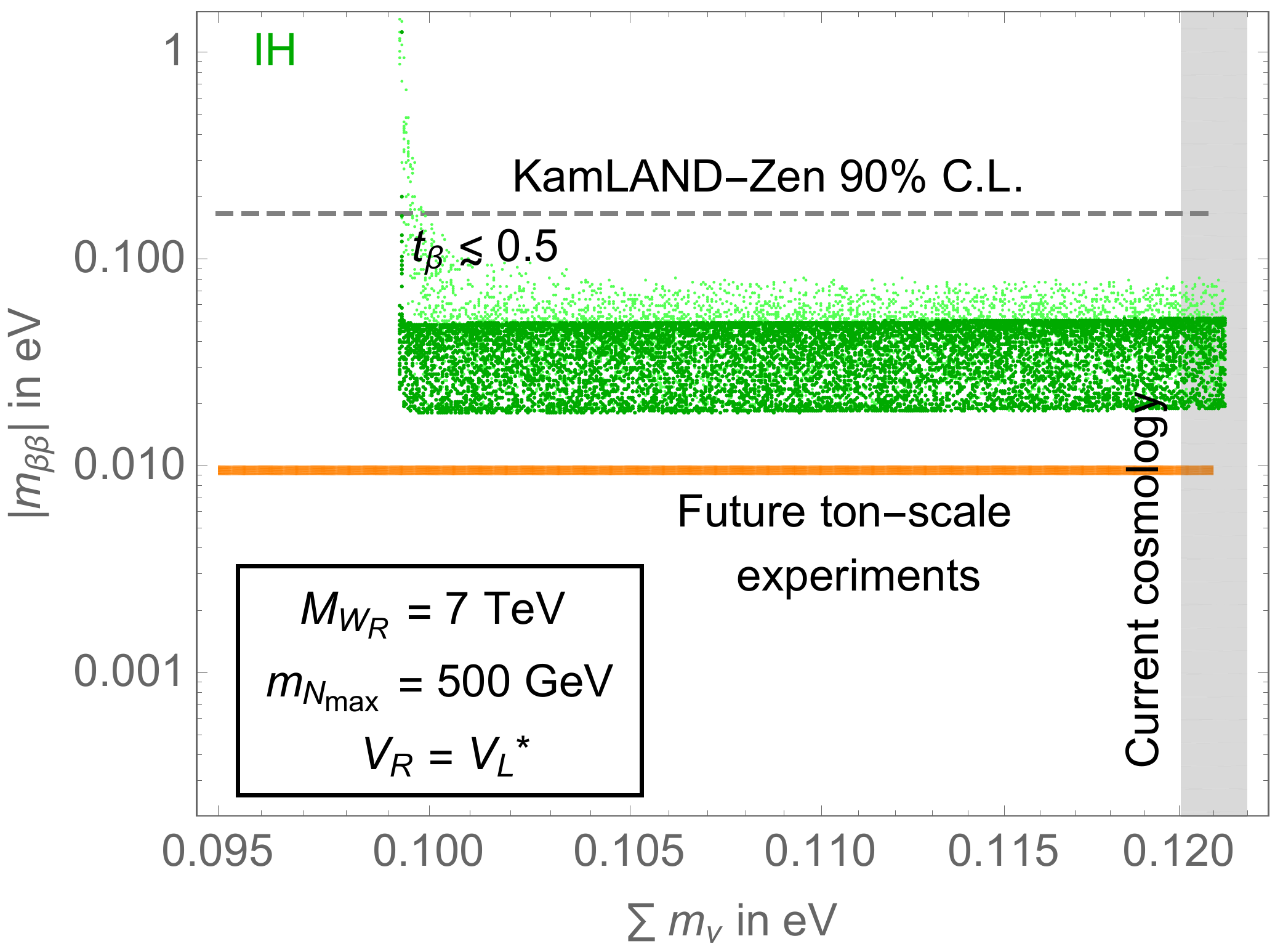}
\caption{Effective Majorana mass as a function of the sum of light neutrino masses. The red and green regions as well as gray and the orange lines have the same meaning of Fig.~\ref{decayrates_mvmin}. The current  constraint from cosmological experiments~\cite{Aghanim:2018eyx} is depicted in the gray region.  }
  \label{decayrates_mvsum}
\end{figure}

Future ton-scale experiments searching for $\onbb$-decay in $^{136}\text{Xe}$ are considered for numerical results. The phase space factor $G_{0\nu}^{-1}=7.11\times 10^{24}~\text{eV}^2\cdot\text{yr}$~\cite{Kotila:2012zza,Stoica:2013lka}, and the nuclear matrix elements (NMEs) $\mathcal{M}_{\nu}=2.91$, $\mathcal{M}_0= -2.64$, $\mathcal{M}_1= -5.52$ and $\mathcal{M}_2= -4.20$, $\mathcal{M}_{NN}=-1.53$ are quoted~\cite{Hyvarinen:2015bda}. We obtain that  $\delta_{\pi\pi}=0.046$, $\delta_{NN\pi}=0.042$, $\delta_{NN}^{\nu}=-0.0096$, and $\delta_{NN}=0.063$, clearly showing the expected chiral suppression $|\mathcal{A}^\mathrm{NNLO}/\mathcal{A}^\mathrm{LO}|\sim 15-20$ or even larger. Again, the LEC $\ell_{\nu}^{NN}\sim \mathcal{O}(1)$ in NDA  and is larger requiring LO $\bar{N}N\bar{N}N\bar{e}e^c$ counterterm~\cite{Cirigliano:2018hja}.

In Fig.~\ref{decayrates_mvmin}, we show the effective Majorana mass $|m_{\beta\beta}|$
as a function of $m_{\nu_{\text{min}}}$ with $m_{N_{\text{max}}}=500$~GeV and $M_{W_R}=7\tev$. To illustrate the impact of the LR contribution, we give the allowed regions with $\tan\beta=0$ (studied in Refs.~\cite{Tello:2010am,Ge:2015yqa}) and $0<\tan\beta\leq 0.5$  in darker and lighter colors, respectively. For most of the $\tan\beta>0$ parameter space, the long-range pion exchange contribution  dominates over other contributions. In Fig.~\ref{decayrates_mvsum}, we plot the $|m_{\beta\beta}|$ as a function of $\sum m_{\nu}$  along with the current upper bound from cosmology experiments~\cite{Aghanim:2018eyx}. 
In particular, we see from Fig.~\ref{decayrates_mvsum} (upper panel) that in the NH, inclusion of the long-range contribution opens up a significant portion of parameter space accessible to ton-scale experiments. Thus,  even if the future CMB and LSS data would exclude the IH~\cite{Brinckmann:2018owf}, there are good prospects of new physics  at the TeV scale giving the dominant contribution to the $\onbb$-decay  rate  in future ton-scale experiments.

\section*{ Acknowledgements }
GL and JCV would like to thank Jordy de Vries for many valuable discussions.  GL thanks Wouter Dekens and Jordy de Vries for fruitful discussions on the half-life calculations regarding Refs.~\cite{Cirigliano:2018yza,Dekens:2020ttz}, Xiao-Dong Ma and Jiang-Hao Yu for the discussions on RGEs. JCV was supported in part under the U.S. Department of Energy contract DE-SC0015376. GL and MJRM were supported in part under U.S. Department of Energy contract DE-SC0011095.  MJRM was also supported in part under National
Science Foundation of China grant No. 19Z103010239.

\bibliography{0nu2betaLR}

\end{document}